\documentclass[11pt]{article}
\usepackage{amssymb}
\usepackage{amsfonts}
\usepackage{graphicx}
\usepackage{amsmath}
\usepackage{hyperref}

\makeatletter \@addtoreset{equation}{section} \makeatother

\newtheorem{theorem}{Theorem}
\newtheorem{lemma}{Lemma}

\newtheorem{proposition}{Proposition}

\def\fracd{\displaystyle\frac}
\def\sumd{\displaystyle\sum}

\def\intd{\displaystyle\int}

\begin{document}

\title{Central limit theorem for fluctuations of linear eigenvalue statistics of
large random graphs}
\author{ M. Shcherbina$^\dagger$
 \qquad B. Tirozzi$^*$\\
$^\dagger$Institute for Low Temperature Physics, Ukr. Ac. Sci \\
$^*$Department of Physics, Rome University "La Sapienza"
 }
\date{}

\maketitle

\begin{abstract} We consider the adjacency matrix $A$ of a large
random graph and study  fluctuations of the function
$f_n(z,u)=\frac{1}{n}\sum_{k=1}^n\exp\{-uG_{kk}(z)\}$ with
$G(z)=(z-iA)^{-1}$. We prove that the moments of  fluctuations
normalized by $n^{-1/2}$ in the limit $n\to\infty$ satisfy the Wick relations for the
Gaussian random variables. This allows us
to prove  central limit theorem for $\hbox{Tr }G(z)$ and then extend the result on the linear eigenvalue
statistics $\hbox{Tr\,}\varphi(A)$ of any  function $\varphi:\mathbb{R}\to\mathbb{R}$ which
 increases, together with its first two derivatives, at infinity
not faster than an exponential. \end{abstract}

\section{Introduction}\label{s:1}

Random graphs appear in different branches of mathematics and physics
(see monographs \cite{B,JLR} and references there in). It is well known that they are closely
connected with the theory of random matrices, since
there is one to one map between  graphs with $n$ vertices and their adjacency matrices (recall that
by the definition the entries $a_{ij}$ of the adjacency matrix are $1$ if the vertices $i$ and $j$ are
connected and $a_{ij}=0$ otherwise). Commonly, the set of $n$ eigenvalues of the adjacency matrix
is referred to as the spectrum of the graph. The limit when the dimension of the matrix $n$ (the number
of the vertexes of the graphs) tends to infinity
 provides a natural approximation for the spectral properties of random graphs.

One of the classes of the
prime reference in the theory of random graphs is the {\it binomial random graph}
originating by P. Erd\H{o}s (see, e.g. \cite{JLR}).
Given a number $p_n\in (0,1)$, this family of graphs ${\mathbf G}(n,p_n)$
is  defined by taking  the set of all graphs
on $n$ vertices as the space of events with  probability
\begin{equation}\label{1.1}
P(G) = p_n^{e(G)} (1-p_n)^{{n \choose 2} - e(G)},
\end{equation}
where $e(G)$ is the number of edges of $G$. Most of the random graphs
studies are devoted to the cases where $p_n\to 0$ as $n\to\infty$.

Ensemble of random symmetric $n\times n$ adjacency matrices
  $A$ corresponding to (\ref{1.1}) can be represented as $A=\{a_{ij}\}_{i,j=1}^n$ with $a_{ii}=0$,
 and  i.i.d.
  \begin{equation}\label{A}
a_{ij}\!=\! \left\{ \begin{array}{ll} 1,&
\textrm{with} \ \textrm{probability } \ p_n ,
\\0,& \textrm{with} \ \textrm{probability} \ 1-p_n ,\\ \end{array}
\right.
\end{equation}
 For any measurable function $f$ we  denote
$\mathbf{E}\{f(A)\}$ the averaging with respect to all random
variables $\{a_{ij}\}_{1\le i<j\le n}$ and
\begin{equation}\label{dVar}
\mathbf{Var}\{f(A)\}:=\mathbf{E}\{|f(A)-\mathbf{E}\{f(A)\}|^2\}.
\end{equation} The normalized eigenvalue counting measure  of  $A$
is defined by the formula
\[ N_n(\lambda)=n^{-1}\sharp
\{j:\lambda^{(n)}_j\!<\!\lambda\}. \]
The ensemble of adjacency
matrices (\ref{A}) is a  particular  case of the random matrix
theory, where the limiting transition $n\to\infty$ is intensively
studied during half of century since the pioneering works by E.
Wigner \cite{W}. Spectral properties of random adjacency matrix
(\ref{A}) were examined in the limit $n\to\infty$ both in numerical
and theoretical physics studies \cite{E1, E2, E3,MF:91,RB:88,RD:90}.
There are  two major asymptotic regimes: $p_n \gg 1/n$ and $p_n =
O(1/n)$ and corresponding models can be called {\it dilute random
matrices} and {\it sparse random matrices}, respectively. The first
studies of spectral properties of sparse and dilute random matrices
in the physical literature are related with the works  \cite{RB:88},
\cite{RD:90}, \cite{MF:91}, where equations for the limiting density
of states of sparse random matrices  were derived. In  papers
\cite{MF:91} and \cite{FM:96} a number of important  results on the
universality of the correlation functions and the Anderson
localization transition were obtained.
  Unfortunately these results were obtained with  the non rigorous  replica and
super symmetry methods.

On mathematical level of rigor the  eigenvalue distribution of dilute
random matrices was studied in \cite{KKPS}. It was shown that
  the normalized eigenvalue counting measure of
$(n p_n)^{-1/2}A$ converges in the limit $n p_n \to \infty$ to the distribution of explicit form known as the
semicircle, or Wigner law \cite{W}.
In the paper \cite{BG2} the adjacency  matrix  of random graphs (\ref{1.1})
with $p_n = p n^{-1}$ was studied.  It was shown that for any $m$ there exist
non random limiting moments $\lim_{n\to\infty}n^{-1}\hbox{Tr }A^m_n$ and these
moments can be found from the system of certain recurrent relations.
The results of \cite{BG2}  was generalized to the case of weighted random graphs in \cite{KSV},
 where the resolvent of the adjacency matrix was studied and
 equations for the Stieltjes transform $g(z)$ of the limiting eigenvalue
 distribution were derived rigorously (note, that the
 same equation for gaussian weights were obtained  in
  \cite{RB:88}, \cite{RD:90}, \cite{MF:91} by using the replica and
  the super symmetry approaches.)
 It was shown in \cite{KSV} that to prove the existence of the limit
 $\lim_{n\to\infty}g_n(z)=g(z)$,
 where $g_{n}(z)$ is the Stieltjes transform of the normalized
counting function $N_n(\lambda)$
\begin{equation}\label{St}
g_{n}(z)=\intd\fracd{d{N}_n(\lambda)}{\lambda-z}
\end{equation}
we need to study the behavior of the function
\begin{equation}
f_n(z,u)=\frac{1}{n}\sum_{k=1}^ne^{-uG_{kk}(z)},
\label{t3.1}\end{equation}
where
\begin{equation}\label{G}
G_{jk}(z)=(z-iA)^{-1}_{jk},
\end{equation}
The function $f_n(z,u)$
 is defined for any $u,z$ such that $\Re z\not=0$.
In what follows it will be important for us that
\begin{eqnarray}\label{f<1}
&&||G||\le |\Re z|^{-1},\quad\sum_{j=1}^n|G_{ij}|^2=(GG^*)_{ii}\le||G||^2\le |\Re z|^{-2}\\
&&\Re (Ge,e)\Re z\ge 0,\;\;\forall e\in\mathbb{R}^n\,\Rightarrow\;\;
|e^{-u(Ge,e)}|\le 1 ,\;\;\hbox{if}\;\;u\,\Re z>0.\notag
\end{eqnarray}
Here and everywhere below $||A||$ means the operator norm of the matrix $A$.

 The following theorem (proven in \cite{KSV}) gives us the limiting properties of
  $f_n(z,u)$ of (\ref{t3.1})
\begin{theorem}\label{t:3} Consider the adjacency matrix (\ref{A}) with $p_n=p/n$.
Then for any $u,z$ such that $u\Re z>0$ we have:

\noindent (i) the variance of the function $f_n(z,u)$ defined by
(\ref{t3.1})
  vanishes in the limit $n\to\infty$:
\begin{equation}
{\bf Var}\{f_n(z,u)\}\le C/(\Re z)^2n,
\label{t3.2}\end{equation}

\noindent (ii) there exists the limit
\begin{equation}
\lim_{n\to\infty}{\bf E}\{f_n(z,u)\}=f(z,u),\quad |{\bf E}\{f_n(z,u)\}-f(z,u)|\le Cu^{1/2}/|\Re z|n^{1/2}
\label{t3.4}\end{equation}

\noindent (iii) if we consider a  class $\mathcal{H}$
 of functions which are
 analytic in $z:\,\,\Re z>0$ and for any fixed $z:\,\,\Re z>0$
 possessing the norm
 \begin{equation}
||f(z)||=\max_{u>0}\frac{|f(z,u)|}{\sqrt{1+u}},
\label{norm}\end{equation}
then  the limiting function is the unique solution in $\mathcal{H}$
 of the functional equation
\begin{equation}
f(z,u)=1-u^{1/2}e^{-p}\int_0^\infty dv
\frac{\mathcal{J}_1(2\sqrt{uv})}{\sqrt v}
\exp\{-zv+pf(z,v)\}, \label{t3.3}\end{equation} where
$\mathcal{J}_1(\zeta)$  is the Bessel function
\begin{equation}
\mathcal{J}_1(\zeta)=\frac{\zeta}{2}\sum_{k=0}^{\infty}
\frac{(-\zeta^2/4)^k}{k!(k+1)!}. \label{J_1}\end{equation}
\end{theorem}

One can  easily see that
$$\begin{array}{l}
-\fracd{\partial}{\partial u}f_n(z,u)\bigg|_{u=0}
=\fracd{1}{n}\sumd_{k=1}^n{\bf E}\{G_{kk}(z)\}=\fracd{1}{n}{\bf E}\{\hbox{Tr }G(z)\}
={\bf E}\{ig_{n}(-iz)\},
\end{array}$$
where $g_{n}(z)$ is the Stieltjes transform (\ref{St}) of the normalized
counting measure ${N}_n(\lambda)$.
Hence, Theorem \ref{t:3} implies that for any $z:\, \Im z\not=
0$
\begin{equation}\label{Varg}
\lim_{n\to\infty}{\bf E}\{|g_{n}(z)-{\bf E}\{g_{n}(z)\}|^2\}=0,
\end{equation}
i.e., the fluctuations of $g_{n}(z)$ vanish in the limit
$n\to\infty$. And (\ref{t3.4}) implies that
\begin{equation}\label{g_p}
g(z)=\lim_{n\to\infty}{\bf E}\{g_{n}(z)\}= -\frac{\partial}{\partial
u}f(z,u)\bigg|_{u=0}
 \end{equation}
 Since the Stieltjes transform  uniquely
  determines the measure,  it follows from Theorem \ref{t:3}   that  there
exists the weak limit $N(\lambda)$ of the normalized counting measure
  $N_n(\lambda)$ and the Stieltjes transform   $g(-iz)$ can be obtained as the
  first derivative of the solution of (\ref{t3.3}).
 Using Theorem \ref{t:3} it is not difficult to obtain
the asymptotic expansions for $g(z)$ with respect to $z^{-k}$.
Since it is well known that the coefficients of this expansion are
the moments of the limiting normalized counting measure of eigenvalues,
we obtain the recurrent
formulas for the moments. Besides, constructing the asymptotic
expansion of $g(z)$ with respect to $p^k$,  it is easy to show
that this expansion is convergent for $p<1$. Since in the case
$a_{ij}=0,1$ the coefficients of this expansion are rational
functions of $z$, we can conclude that the limiting spectrum is
pure point and consists of the spectra of finite blocks only.

\medskip

Results of \cite{KSV} described above can be viewed as the analogs of the Law
of Large Numbers for linear eigenvalue statistics
\begin{equation}\label{linst}
    \mathcal{N}_n[\varphi]=\sum_{i=1}^n\varphi(\lambda_i)=\hbox{Tr }\varphi(A)
\end{equation}
corresponding to continuous test functions. Indeed, it follows from
(\ref{Varg}) -- (\ref{g_p}) that for any continuous test function
there exists
\[\lim_{n\to\infty}n^{-1}\mathcal{N}_n[\varphi]=\int \varphi(\lambda)dN(\lambda),\]
where $N$ is the limiting normalized counting measure of eigenvalues.
In the present paper we consider
the central limit theorem, the second element of the standard
 probabilistic analysis of  linear statistics. Similar questions for other ensembles
 of random matrices were studied in \cite{An-Ze:06,Ba-Si:06,Gu:02,Jo:98a,LP:09,Si-So:98a,So:00c}.
Note, however, that for almost all ensembles studied in the
random matrix theory,  like the Wigner ensemble, the
Marchenko-Pastur ensemble, the matrix models, etc the variance of
linear statistics for smooth functions is bounded (see
\cite{An-Ze:06,Ba-Si:06,Gu:02,Jo:98a,LP:09,Si-So:98a,So:00c}). Thus,
for these ensembles, one expects the Central Limit Theorem to be
valid for statistics themselves, i.e., without an  $n$-dependent
normalization factor in front. This has to be compared with the case
of  i.i.d. random variables with finite second moment, where the
variance of linear statistics is always of the order $O(n)$,
$n\to\infty$
 and the Central Limit Theorem is valid for linear statistics
divided by $n^{1/2}$. As we will see below this is the case also for the ensemble of  sparse
adjacency matrices (\ref{A}) with $p_n=p/n$.

The aim of  the present paper is to study the fluctuations of linear eigenvalue statistics for
different classes of  test functions. Following the method of \cite{KSV} we study first the functions
$f_n(z,u)$ (defined in (\ref{t3.1}))
and prove that its fluctuations converges in distribution to the complex Gaussian random variables.

Define the $m$-th generalized moment of the fluctuations of $f_n(z,u)$:
\begin{eqnarray}
M_{m,n}(z_1,u_1;\dots;z_{m},u_m)&:=&n^{-m/2}\mathbf{E}\left\{\prod_{j=1}^m\left(\sum_{k=1}^n
\overset{\circ}e^{-u_jG_{kk}(z_j)}\right)\right\}\notag\\
&=&n^{m/2}\mathbf{E}\left\{\prod_{j=1}^m\overset{\circ}{f_n}(z_j, u_j)\right\},\quad \Re z_i\not=0.
\label{M_1}
\end{eqnarray}
Here and below for any random variable $\xi$ we denote
\[\overset\circ{\xi}=\xi-\mathbf{E}\{\xi\}\]

\begin{theorem}\label{t:1} Consider the adjacency matrix (\ref{A}) with $p_n=p/n$.
Let $M_{m,n}(z_1,u_1;\dots;z_{m},u_m)$ ($m=2,3,\dots$ ) of (\ref{M_1}) be the "moments" of the
fluctuations of $f_n(z,u)$ of (\ref{t3.1}). Then for any $m>2$ and  $z_1,\dots,z_m:\Re z_j>0$
there exists
\begin{equation}\label{t1.1}
M_{m}(z_1,u_1;\dots;z_{m},u_m):=\lim_{n\to\infty}M_{m,n}(z_1,u_1;\dots;z_{m},u_m).
\end{equation}
Moreover, the following recursion equations hold:
\begin{multline}\label{t1.2}
M_{m}(z_1,u_1;\dots;z_{m},u_m)\\=\sum_{j=2}^mM_{2}(z_1,u_1;z_{j},u_j)
M_{m-2}(z_2,u_2;\dots;z_{j-1},u_{j-1};z_{j+1},u_{j+1};\dots;z_{m},u_m).
\end{multline}
\end{theorem}

Theorem \ref{t:1} can be used
to prove the central limit theorem for  fluctuations of the trace of
$G(z)$ of (\ref{G}). Indeed, if we denote
\begin{equation}\label{M^*}
M_{m,n}^*(z_1,\dots,z_{m}):=n^{-m/2}
\mathbf{E}\left\{\hbox{Tr }\overset\circ{G}(z_1)\dots\hbox{Tr }\overset\circ{G}(z_{m})\right\},
\end{equation}
then it is easy to see that
\[M_{m,n}^*(z_1,\dots,z_{m})=\frac{\partial^m}{\partial u_1\dots \partial u_m}
M_{m}(z_1,u_1;\dots;z_{m},u_m)\bigg|_{u_1=\dots=u_m=0}.
\]
Since $M_{m}(z_1,u_1;\dots;z_{m},u_m)$ are evidently analytic in each $u_i$  in
 some neighborhood of $u_i=0$ and  bounded uniformly
in $n$ for any fixed $z_1,\dots,z_m$ ($\Re z_i\not=0$) (see Lemma \ref{l:1} below),
 we pass to the limit $n\to\infty$ in the above relations  and obtain the following
theorem:
\begin{theorem}\label{t:2} Let $G(z)$ be the resolvent (\ref{G}) of the sparse adjacency
matrix (\ref{A}) with $p_n=p/n$. Then for any $m>2$ and  $z_1,\dots,z_m:\Re z_j>0$
there exists
\begin{equation}\label{t2.1}
M_{m}^*(z_1,\dots,z_{m}):=\lim_{n\to\infty}M_{m,n}^*(z_1;\dots;z_{m})
\end{equation}
and the following recursions hold:
\begin{equation}\label{t2.2}
M_{m}^*(z_1,\dots,z_{m})=\sum_{j=2}^m M_{2}^*(z_1,z_{j})
M_{m-2}^*(z_2,\dots,z_{j-1},z_{j+1},\dots,z_{m})
\end{equation}
\end{theorem}
Theorem \ref{t:2} by a standard way implies the central limit
theorem for $v_n(z)=n^{-1/2}\hbox{Tr }\overset\circ{ G}(z)$. Indeed, if we put in  (\ref{t2.1}) -- (\ref{t2.2})
$z_1=z_2=...=z_m=z$, then Theorem \ref{t:2} yields that there exist limits of all moments
of the complex random variable $v_n(z)$ and
\[M_{2m}(z):=\lim\mathbf{E}\{(n^{-1/2}\hbox{Tr }\overset\circ{ G}(z))^{2m}\}=(2m-1)!!(M_2(z))^m\]
This means that $v_n(z)$ converges in distribution to a complex Gaussian random variable with
zero mean and  variance $M_2(z)$.

It is possible also to derive   from Theorem \ref{t:2} the central limit theorem for
the linear eigenvalue statistics of any  function $\varphi$ which  grows not
faster than  an exponent at infinity and possesses two
derivatives  with the same property, i.e. there exists a constant $c>0$ such that
$\varphi,\,\varphi',\,\varphi''\in
L^2(\mathbb{R},\cosh^{-2}(c\lambda))$. Here and below
\begin{equation}\label{L_2}
L^2(\mathbb{R},w(\lambda))=\left\{f:\int_{\mathbb{R}}|f(\lambda)|^2w(\lambda)d\lambda<\infty\right\}
\end{equation}
\begin{theorem}\label{t:4} Consider the adjacency matrix (\ref{A}) with $p_n=p/n$ and
take any  function $\varphi$ which  possesses two
derivatives   such that $\varphi,\,\varphi',\,\varphi''\in
L^2(\mathbb{R},\cosh^{-2}(c\lambda))$ with some constant $c>0$. Then the random variable
$n^{-1/2}\overset\circ{\mathcal{N}_n}[\varphi]$ converges in distribution to  a Gaussian
random variable with zero mean and variance
$V[\varphi]:=\displaystyle\lim_{n\to\infty}\mathbf{Var} \{n^{-1/2}\mathcal{N}_n[\varphi]\}$.
\end{theorem}

 It is clear from the above discussion  that Theorem \ref{t:1} plays a key role in the paper,
because Theorems \ref{t:2} and \ref{t:4} are in fact corollaries
of Theorem \ref{t:1}.
The proof of Theorem \ref{t:1} is based on a version of the cavity method which has been
 used many times for proving different limiting relations of statistical mechanics and random matrices.
 The idea is to compare the behavior of the object function (e.g. free-energy, resolvent, etc.)
 for the complete system of random variables of the problem and the one with some subset of random variables replaced by 0.

\medskip

Let us try to explain the connections among the lemmas and  propositions which are necessary for the proof
of Theorem \ref{t:1}.  The proof should be seen as a logical sequence
 of the following steps:

 \begin{itemize}

 \item  We prove first   bounds on $M_{m,n}(z_1,u_1;\dots;z_{m},u_m)$ uniform  in $ (z_1,u_1;\dots;z_{m},u_m)$
 ($\Re z_j\ge C>0$)  (see Lemma \ref{l:1}).  One uses the norm estimates of the martingale theory
 (Proposition \ref {l:1}), identities
 for the resolvent and the cavity method  consisting in studying the difference of the resolvent of the full
 matrix and the same matrix without the first line and the first column.

 \item  To prove the convergence of the variance of the sums of  exponentials we
  need to generalize  Theorem \ref{t:3} and to show the existence of the limits of
 exponentials multiplied by some entire functions (cf Lemma \ref{l:3} and Theorem \ref{t:3}). The
 proof of Lemma \ref{l:3} is based on the relations for  some functions of $A$
 given by Proposition \ref{p:2}.

 \item Lemma \ref{l:2} proves the self averaging properties and the existence of the limits for the
 terms which will appear in the proof of CLT.

 \item Finally we prove that the "moments" (\ref{M_1}) as functions of $u_i$ satisfy the linear integral
 equations with the kernel defined in terms of the function $f$ of (\ref{t3.3}) (see (\ref{t1.15})).
 Since we are able  to prove that these equations are uniquely solvable for $\Re z>M_0$ with some
 fixed $M_0$, we finish the proof of Theorem \ref{t:1}.

 \end{itemize}

\section{Proofs}

 We start from the  lemma which gives bounds for $M_{m,n}$.

\begin{lemma}\label{l:1} For any $m\in\mathbb{N}$ and  $z_1,\dots,z_m:\Re z_j>0$
  there exists a constant $C_m$ such that uniformly in $u_1\dots,u_m>0$
  \begin{equation}\label{b_M}
|M_{m,n}(z_1,u_1;\dots;z_{m},u_m)|\le C_m
\end{equation}
\end{lemma}

 The proof is based on the martingale property of the sequence of averages of the
 functions of the random matrix $A$ with respect to its rows or columns.
 The sequence is ordered with respect to the index of the rows and the proposition
 below is based on the sequence of the conditional expectations like in the proof of
 self-averaging of the free-energy for  disordered systems.

 \begin{proposition}
\label{p:mart} Let $\xi _{\alpha },\;\alpha =1,...,\nu $ be independent
random variables, assuming values in $\mathbb{R}^{m_{\alpha }}$ and having
probability laws $P_{\alpha }$, $\alpha =1,\dots ,\nu $ and let $\Phi :%
\mathbb{R}^{m_{1}}\times \dots \times \mathbb{R}^{m_{\nu }}\rightarrow
\mathbb{C}$ be a Borelian function. Set
\begin{equation}
\Phi _{\alpha }(\xi _{1},\dots ,\xi _{\alpha })=\int \Phi (\xi _{1},\dots
,\xi _{\alpha },\xi _{\alpha +1},\dots ,\xi _{\nu })P_{\alpha +1}(d\xi
_{\alpha +1})\dots P_{\nu }(d\xi _{\nu })  \label{phi_j}
\end{equation}%
so that $\Phi _{\nu }=\Phi ,\quad \Phi _{0}=\mathbf{E}\{\Phi \}$,
where $\mathbf{E}\{\dots \}$ denotes the expectation with respect to the
product measure $P_{1}\dots P_{\nu }$.

Then for any positive $p\geq 1$ there exists $C_{p}^{\prime }$, independent
of $\nu $ and such that
\begin{equation}
\mathbf{E}\{|\Phi -\mathbf{E}\{\Phi \}|^{2p}\}\leq C_{p}^{\prime }\nu
^{p-1}\sum_{\alpha =1}^{\nu }\mathbf{E}\{|\Phi _{\alpha }-\Phi _{\alpha
-1}|^{2p}\}.  \label{martk}
\end{equation}
Moreover, if
for every $\alpha =1,\dots ,\nu $ there exists a $\xi _{\alpha }$%
-independent $\Psi ^{(\alpha )}:\mathbb{R}^{m_{1}}\times \dots \times
\mathbb{R}^{m_{\nu }}\rightarrow \mathbb{C}$ such that%
\begin{equation}\label{ppcg}
\mathbf{E}\{|\Phi -\Psi ^{(\alpha )}|^{2p}\}\le C<\infty ,\quad \alpha =1,\dots
,\nu ,
\end{equation}%
then
\begin{equation}\label{ECq}
\mathbf{E}\{|\Phi -\mathbf{E}\{\Phi \}|^{2p}\}\leq 2C_{p}^{\prime
}C\nu ^{p}.
\end{equation}
\end{proposition}

\textit{Proof.} The proof of (\ref{martk}) is given in
 \cite{Dh-Co:68}. Hence, we show only how to derive (\ref{ppcg}) from (\ref{martk}).
It follows from (\ref{phi_j}) and (\ref{ppcg}) that the integrals of $\Psi
^{(\alpha )}$ with respect to $P_{\alpha +1}...P_{\nu }$ and $P_{\alpha
}P_{\alpha +1}...P_{\nu }$ \ coincide and we obtain
\begin{eqnarray*}
\mathbf{E}\{|\Phi _{\alpha }-\Phi _{\alpha -1}|^{2p}\} &\leq &2^{2p-1}(%
\mathbf{E}\{|(\Phi -\Psi ^{(\alpha )})_{\alpha -1}|^{2p}\}+\mathbf{E}%
\{|(\Phi -\Psi ^{(\alpha )})_{\alpha }|^{2p}\}) \\
&\leq &2^{2p}\mathbf{E}\{|\Phi -\Psi ^{(\alpha )}|^{2p}\}.
\end{eqnarray*}%
This and (\ref{martk}) prove (\ref{ECq}).
$\square$

\textit{Proof of Lemma \ref{l:1}} The H\"{o}lder inequality yields
\[|M_{m,n}(z_1,u_1;\dots;z_{m},u_m)|\le n^{m/2}\prod_{j=1}^m\mathbf{E}\bigg\{\bigg|
\overset{\circ}f_n(z_j,u_j)\bigg|^m\bigg\}^{1/m}.\]
Hence,  it suffices to prove the bound for the r.h.s. of the above inequality.
 For this   we use Proposition \ref{p:mart}
for  the function $\Phi=nf_n(z,u)$ with $f_n(z,u)$ of (\ref{t3.1}).
  According to (\ref{ECq}) and the  approach of the cavity method for our purposes
it is enough to choose the functions $\Psi_i$ independent from $\xi_i=a^{(i)}:=
(a_{i1},\dots ,a_{ii-1},0,a_{ii+1},\dots,a_{in})$ and prove (\ref{ppcg}).
Set
\begin{eqnarray}\label{A^1}
A^{(i)}&=&A\bigg|_{a_{ij}=0,j=1,\dots,n},\quad G^{(i)}(z)=(z-iA^{(i)})^{-1},\\
\Psi^{(i)}&=&nf^{(i)}_n(z,u):=\sum_{k\not=i}
e^{-uG_{kk}^{(i)}(z)}.
\label{f^1}
\end{eqnarray}
By the symmetry reason it suffices to prove (\ref{ppcg}) for $i=1$.
We use the representations:
\begin{eqnarray}
G_{ij}(z)&=&G_{ij}^{(1)}(z)-\frac{(G^{(1)}a^{(1)})_i(G^{(1)}a^{(1)})_j}{z+
(G^{(1)}a^{(1)},a^{(1)})},\quad i,j\not=1,\notag\\
G_{1j}(z)&=&\frac{i(G^{(1)}a^{(1)})_j}{z+
(G^{(1)}a^{(1)},a^{(1)})},\quad j\not=1,\label{rep_1}\\
G_{11}(z)&=&(z+
(G^{(1)}a^{(1)},a^{(1)}))^{-1},
\notag\end{eqnarray}
where $a^{(1)}=(0,a_{12},\dots,a_{1n})$. The  inequality
$|e^x-e^y|\le |x-y|\max\{|e^x|,|e^y|\}$ and (\ref{f<1}) imply
\begin{eqnarray}\label{l1.3}
\left|\sum_{k=2}\left(e^{-uG_{kk}}-e^{-uG^{(1)}_{kk}}\right)\right|
\le u\sum_{k=2}\left|G_{kk}-G^{(1)}_{kk}\right|\hskip4.5cm\\
\le u\sum_{k}\frac{(G^{(1)}a^{(1)})_k\overline{(G^{(1)}a^{(1)})_k}}{|z+
(G^{(1)}a^{(1)},a^{(1)})|}
= u
\frac{(G^{(1)}(z)a^{(1)},G^{(1)}( z)a^{(1)})}{|z+
(G^{(1)}a^{(1)},a^{(1)})|}.\notag
\end{eqnarray}
But the spectral theorem yields
\[(G^{(1)}(z)a^{(1)},G^{(1)}( z)a^{(1)})=\sum_{j=1}^n\frac{|(\psi^{(j)},a^{(1)})|^2}{(\lambda^{(j)}-\Im z)^2+
(\Re z)^2}= \frac{1}{\Re z}\Re(G^{(1)}a^{(1)},a^{(1)}),\]
where $A^{(1)}\psi^{(j)}=\lambda^{(j)}\psi^{(j)}$. Thus, since by (\ref{f<1}) $\Re z\,\Re(G^{(1)}a^{(1)},a^{(1)})>0$, we
have
\begin{equation}\label{ineq*}
\frac{(G^{(1)}(z)a^{(1)},G^{(1)}( z)a^{(1)})}{|z+
(G^{(1)}a^{(1)},a^{(1)})|}\le(\Re z)^{-1}.
\end{equation}
 Inequality (\ref{ppcg})
 for our choice of  $\Phi$ and $\Psi^{(i)}$   follows from (\ref{l1.3}) and (\ref{ineq*}).
$\square$

\medskip

 In the proof of Theorem \ref{t:1} we will replace sometimes $M_{m,n}$
by the moments independent of $\{a_{1,j}\}_{j=2}^n$.
Set
\begin{eqnarray}
M_{m,n}^{(1)}(z_1,u_1;\dots;z_{m},u_m)&:=&n^{m/2}\mathbf{E}\left\{\prod_{j=1}^m
\overset{\circ}f_n^{(1)}(z_j,u_j)\right\}
\label{M_1^1}
\end{eqnarray}
with $f_n^{(1)}$ of (\ref{f^1}). Note that (\ref{l1.3}) yields that
for any $m\in\mathbb{N}$ and  $z_1,\dots,z_m:\Re z_j>0$
  there exists  constants $C_m, C_m'$ such that uniformly in $u_1\dots,u_m$
  \begin{eqnarray}
&&|M_{m,n}(z_1,u_1;\dots;z_{m},u_m)-M_{m,n}^{(1)}(z_1,u_1;\dots;z_{m},u_m)|\le C_m n^{-1/2},\notag\\
&&|M_{m,n}^{(1)}(z_1,u_1;\dots;z_{m},u_m)|\le C_m'.
\label{M^1-M}\end{eqnarray}

To study the behavior of some functions,
depending on $\{a_{1,j}\}_{j=2}^n$, we use the proposition:
\begin{proposition}\label{p:2}
Let $\mathbf{E}_1\{\dots\}$ be the averaging with respect to
$\{a_{1k}\}_{k=2}^n$. Then  we have for any $u,v>0$ and $\Re z>2$
\begin{eqnarray}\label{p2.*}
&&e^{-v(G^{(1)}a^{(1)},a^{(1)})}=e^{-v\sum_{k}G^{(1)}_{kk}a_{1k}}+r_{v},
\\   r_{v}&=&v\sum_{i\not=j}G^{(1)}_{ij}a_{1i}a_{1j}
+O\bigg(v^2\bigg|\sum_{i\not=j}G^{(1)}_{ij}a_{1i}a_{1j}\bigg|^2\bigg),
\quad\mathbf{E}_1^{1/2}\{|r_{v}|^2\}\le Cvn^{-1/2}.
\notag\end{eqnarray}
 Moreover, denoting $Z=z+(G^{(1)}a^{(1)},a^{(1)})$, we have
\begin{eqnarray}\label{p2.0}
\mathbf{E}_1\bigg\{\sum_{j=2}^n(e^{-uG_{jj}}-e^{-uG_{jj}^{(1)}})\bigg\}
&=&\mathbf{E}_1\bigg\{\sum_{j,k=2}^ne^{-uG_{jj}^{(1)}}(e^{u(G^{(1)}_{jk})^2/Z}-1)a_{1,k}\bigg\}\\
&&+ O\bigg(\frac{(u+u^2)e^{u}}{n}\bigg)
\notag\end{eqnarray}
\end{proposition}
\textit{Proof}. Note that since  $|\Re z|^{-1}\le 1/2$,
everywhere below we will replace $|\Re z|^{-1}$ by a constant.
We need below the trivial bounds:
\begin{equation}\label{d_exp}
|e^{a}-e^{b}|\le|a-b|\max\{|e^{a}|,|e^{b}|\},\quad
|e^{a}-e^{b}-(a-b)|\le|a-b|^2\max\{|e^{a}|,|e^{b}|\}.
\end{equation}
The first bound  and the second line of (\ref{f<1}) combined with (\ref{rep_1}) imply
\[|e^{-v(G^{(1)}a^{(1)},a^{(1)})}-e^{-v\sum_{k}G^{(1)}_{kk}a_{1k}}|
\le v\bigg|\sum_{k_1\not=k_2}G^{(1)}_{k_1k_2}a_{1k_1}a_{1k_2}\bigg|\]
Averaging the square of the bound we obtain
\begin{eqnarray}\label{p2.**}
&&\mathbf{E}_1\{|r_v|^2\}=v^2\mathbf{E}_1\bigg\{\sum_{k_1\not=k_2,k_3\not=k_4}G^{(1)}_{k_1k_2}
\overline{G^{(1)}_{k_3k_4}} a_{1k_1}a_{1k_2}a_{1k_3}a_{1k_4}\bigg\}\\
&\le& \frac{C_1v^2}{n^2}\sum_{k_1,k_2}|G^{(1)}_{k_1k_2}|^2+
\frac{C_2v^2}{n^3} \sum_{k_1,k_2,k_3}G_{k_1k_2}^{(1)}\overline{G^{(1)}_{k_1k_3}}+\frac{C_3v^2}{n^4}
\bigg|\sum_{k_1,k_2}G_{k_1k_2}^{(1)}\bigg|^2\le\frac{C_4v^2}{n}.
\notag\end{eqnarray}
Here we used  the bounds valid for any matrix $A$:
\begin{equation}\label{p2.a}
\bigg|\sum_{j,k}A_{jk}\bigg|\le n||A||,\quad
\sum_{k}|A_{jk}|\le n^{1/2}\bigg(\sum_{k}|A_{jk}|^2\bigg)^{1/2}\le n^{1/2}||A||.
\end{equation}
To prove  (\ref{p2.0}) we show first that
\begin{eqnarray}\notag
  &&  \bigg|\sum_{j=2}^n\mathbf{E}_1\{(\exp\{-uG_{jj}\}-\exp\{-uG_{jj}^{(1)}+\sum
(G^{(1)}_{jk})^2a_{1k}/Z\}\bigg|\\
&&:=\bigg|\sum_{j=2}^n\mathbf{E}_1\{e^{a_j}-e^{b_j}\}\bigg|\le Cn^{-1/2}e^{u}.
\label{l2.1}\end{eqnarray}
The second inequality of (\ref{d_exp}) and the bounds that $|e^{a_j}|\le 1$ and $|e^{b_j}|\le e^{u/|\Re z|^3}\le e^u$
yield
\begin{eqnarray*}\notag
&&\bigg|\sum_{j=2}^n\mathbf{E}_1\{e^{a_j}-e^{b_j}\}\bigg|
\le\bigg|\sum_{j=2}^n\mathbf{E}_1\{a_j-b_j\}\bigg|
+\sum_{j=2}^n\mathbf{E}_1\left\{\left|(e^{a_j}-e^{b_j})-(a_j-b_j)\right|\right\}\\
&&\le\bigg|\sum_{j=2}^n\mathbf{E}_1\{a_j-b_j\}\bigg|+
e^{u}\sum_{j=2}^n\mathbf{E}_1\left\{|a_j-b_j|^2\right\}
\end{eqnarray*}
Then, similarly to (\ref{p2.**}) we have
\begin{eqnarray*}
\bigg|\sum_{j=2}^n\mathbf{E}_1\{a_j-b_j\}\bigg|&=&u\bigg|\sum_{j=2}^n\mathbf{E}_1\bigg\{
\sum_{k_1\not=k_2}G_{jk_1}^{(1)}G_{jk_2}^{(1)}a_{1k_1}a_{1k_2} Z^{-1}\bigg\}\bigg|\\
&\le&u\mathbf{E}_1^{1/2}\bigg\{\bigg|\sum_{k_1\not=k_2}(G^{(1)}G^{(1)})_{k_1,k_2}a_{1k_1}a_{1k_2}\bigg|^2\bigg\}
\mathbf{E}_1^{1/2}\{Z^{-2}\}\le Cun^{-1/2}.
\end{eqnarray*}
Here we  used also that in view of (\ref{f<1}) $\Re Z>\Re z\ge 2$. Moreover,  similarly to
(\ref{p2.**}), we obtain
\begin{eqnarray*}
\mathbf{E}_1\left\{|a_j-b_j|^2\right\}\le u^2\mathbf{E}_1\bigg\{\bigg|\sum_{k_1\not=k_2,k_3\not=k_4}
G_{jk_1}^{(1)}G_{jk_2}^{(1)}a_{1k_1}a_{1k_2}\overline{G_{jk_3}^{(1)}}\overline{G_{jk_4}^{(1)}}
a_{1k_3}a_{1k_4}\bigg|\bigg\}
\le Cu^2n^{-2}.
\end{eqnarray*}
Summing with respect to $j$, we get (\ref{l2.1}).
Besides, we have
\begin{eqnarray*}
&&\sum_{j=2}^n\left(\exp\{-uG_{jj}^{(1)}+\sum
(G^{(1)}_{jk})^2a_{1k}/Z\}-\exp\{-uG_{jj}^{(1)}\}\right)\\
&&=\sum_{j=2}^ne^{-uG_{jj}^{(1)}}
\sum_{m=1}^\infty\frac{u^m(\sum_k(G^{(1)}_{jk})^2a_{1k})^m}{m!Z^m}\\
&&=\sum_{j,k=2}^ne^{-uG_{jj}^{(1)}}
\sum_{m=1}^\infty\frac{u^m(G^{(1)}_{jk})^{2m}a_{1k}}{m!Z^m}+r_n,
\end{eqnarray*}
where the remainder term $r_n$ admits the bound
\begin{eqnarray*}
\mathbf{E}_1\{|r_n|\}\le\sum_{j=2}^n\sum_{k_1\not=k_2}
\mathbf{E}_1\left\{|G^{(1)}_{jk_1}|^2|G^{(1)}_{jk_2}|^2a_{1k_1}a_{1k_2}\right\}
\sum_{m=2}^\infty\frac{u^m(\sum_k|G^{(1)}_{jk}|^{2})^{m-2}}{2(m-2)!Z^m}\le \frac{Cu^2e^{u}}{n}.
\end{eqnarray*}
The averaging here is similar to (\ref{p2.**}). Thus, we have proved (\ref{p2.0})$\square$

\medskip

Set (cf (\ref{J_1}))
\begin{equation}\label{ti-J}
    \widetilde{\mathcal{J}}_1(\zeta)=\sum_{k=0}^\infty\frac{\zeta^{k+1}}{k!(k+1)!}
    =-2i\zeta^{1/2}\mathcal{J}_1(2i\zeta^{1/2})
\end{equation}
Below we will need the following properties of $\widetilde{\mathcal{J}}_1(\zeta)$
\begin{equation}\label{pr_J}
\sup_{|\zeta|\le r}|\widetilde{\mathcal{J}}_1(\zeta)|\le \widetilde{\mathcal{J}}_1(r),\quad
|\widetilde{\mathcal{J}}_1(\zeta)|\le |\zeta|(1+\widetilde{\mathcal{J}}_1(|\zeta|)),\quad
\sup_{|\zeta|\le r}|\widetilde{\mathcal{J}}_1''(\zeta)|\le(1+\widetilde{\mathcal{J}}_1(r))
\end{equation}
The following lemma is the analog of Theorem \ref{t:3} for the function
which will appear in the proof of Theorem \ref{t:2}.
\begin{lemma}\label{l:3}
 For any $u>0$, $v\in\mathbb{C}$, $\Re z>2$, and $\mathcal{J}_1$ of (\ref{J_1}
the random variable
\begin{equation}\label{V_F}
    V_{J,n}(z,u,v)=n^{-1}\sum_{j,k=1}^ne^{-uG_{kk}}\widetilde{\mathcal{J}}_1(vG_{kj}^2)
\end{equation}
possesses the property:
\begin{equation}\label{s-a.2}
    \mathbf{Var}\{V_{J,n}(z,u,v)\}\le n^{-1}q(u,|v|)(1+\widetilde{\mathcal{J}}_1^2(|v|))
\end{equation}
where $q(u,v)$ is a fixed polynomial. Moreover,  there exists
\begin{equation}\label{p1.1}
V_{J}(z,u,v):=\lim_{n\to\infty}\mathbf{E}\{V_{J,n}(z,u,v)\}.
\end{equation}
and
\begin{equation}\label{p1.1a}
|r_{J,n}(z,u,v)|:=|V_{J,n}(z,u,v)-V_{J}(z,u,v)|\le Cn^{-1/2}(1+\widetilde{\mathcal{J}}_1(|v|)).
\end{equation}
\end{lemma}
\textit{Proof}.
According to Proposition \ref{p:mart} to prove (\ref{s-a.2}) it is enough to prove that
\begin{equation}\label{p1.2}
\Delta^{(1)}:=\bigg|\sum_{j,k}\left(e^{-uG_{kk}}\widetilde{\mathcal{J}}_1(vG_{kj}^2)
-e^{-uG^{(1)}_{kk}}\widetilde{\mathcal{J}}_1(v(G^{(1)}_{kj})^2)
\right)\bigg|\le q_1(u,v)(1+\widetilde{\mathcal{J}}_1(|v|))
\end{equation}
with polynomial $q_1$. Then $q=q_1^2$.
  In view of the second bound of (\ref{pr_J}), (\ref{f<1}), (\ref{rep_1}), (\ref{l1.3}),  and (\ref{ineq*}) we have
\begin{eqnarray*}\notag
\Delta^{(1,1)}&:=&\bigg|\sum_{j,k}\left(e^{-uG_{kk}}-e^{-uG^{(1)}_{kk}}\right)
\widetilde{\mathcal{J}}_1(vG_{kj}^2)\bigg|\\&&
\le Cu|v|(1+\widetilde{\mathcal{J}}_1(|v|))\sum_{j,k}\left|G_{kk}-G^{(1)}_{kk}\right||G_{kj}^2|
\\&\le& Cu|v|\widetilde{\mathcal{J}}_1(|v|)
\frac{(G^{(1)}(z)a^{(1)},G^{(1)}( z)a^{(1)})}{|z+
(G^{(1)}a^{(1)},a^{(1)})|}\le Cu|v|(1+\widetilde{\mathcal{J}}_1(|v|)).\notag
\end{eqnarray*}
Moreover, by the third bound of (\ref{pr_J}), we can write
\begin{eqnarray}\notag
\Delta^{(1,2)}&:=&\bigg|\sum_{j,k}e^{-uG^{(1)}_{kk}}\left(\widetilde{\mathcal{J}}_1(vG_{kj}^2)
-\widetilde{\mathcal{J}}_1(v(G^{(1)}_{kj})^2)\right)\bigg|
\le \bigg|v\sum_{j,k}e^{-uG^{(1)}_{kk}}\left(G_{kj}^2-G^{(1)2}_{kj}\right)\bigg|\\&&+
C|v|^2(1+\widetilde{\mathcal{J}}_1(|v|))\sum_{j,k}|G_{kj}-G^{(1)}_{kj}|(|G_{kj}|^3+|G^{(1)}_{kj}|^3)
\label{p1.4}\end{eqnarray}
Then denoting $\Sigma_1$ the first sum in the r.h.s., we have in view of the first line of
(\ref{rep_1}), (\ref{f<1}),  and (\ref{ineq*}),:
\begin{eqnarray*}
|\Sigma_1|&=&\bigg|\sum_{j,k}\frac{(G^{(1)}(z)a^{(1)})_k(G^{(1)}(z)a^{(1)})_j}{|z+
(G^{(1)}a^{(1)},a^{(1)})|}
\left(G_{jk}(z)+G^{(1)}_{jk}(z)\right)\bigg|\\
&\le &(||G(z)||+||G^{(1)}(z)||)\frac{(G^{(1)}(z)a^{(1)},G^{(1)}(z)a^{(1)})}{|z+
(G^{(1)}a^{(1)},a^{(1)})|}\le C.
\end{eqnarray*}
To estimate $\Sigma_2$ -- the second  sum  in the r.h.s. of (\ref{p1.4}) we use that
 for any matrix $M$ if we consider the matrix
$M^{(2)}=\{|M|_{i,j}^2\}_{i,j=1}^n$, then
\begin{equation}\label{M^2}
||M^{(2)}||\le\sup_{i}\bigg(\sum_j|M_{ij}|^2\bigg)^{1/2}\sup_{j}\bigg(\sum_i|M_{ij}|^2\bigg)^{1/2}\le
||M||^2.
\end{equation}
Hence, the matrix with entries $|G_{kj}|^2$ has the norm bounded by $||G||^2\le|\Re z|^{-2}$.
Then (\ref{M^2}) and (\ref{ineq*}) imply
for $\Sigma_2$:
\begin{eqnarray*}
\Sigma_2&\le & |\Re
z|^{-1}\sum_{j,k}\frac{|(G^{(1)}(z)a^{(1)})_k|\,|(G^{(1)}(z)a^{(1)})_j|}{|z+
(G^{(1)}a^{(1)},a^{(1)})|}
\left(|G_{jk}(z)|^2+|G^{(1)}_{jk}(z)|^2\right)          \\
&\le&2|\Re z|^{-3}\frac{(G^{(1)}(z)a^{(1)},G^{(1)}(z)a^{(1)})}{|z+
(G^{(1)}a^{(1)},a^{(1)})|}\le C.
\notag
\end{eqnarray*}
Thus, we have proved (\ref{p1.2}) and so (\ref{s-a.2}).

To prove (\ref{p1.1}) -- (\ref{p1.1a}) it suffices to prove  that for any $m\ge 2$ there exists
\begin{equation*}
V_m(u,z)=\lim_{n\to\infty}\mathbf{E}\bigg\{n^{-1}\sum_{j,k}e^{-uG_{kk}}G_{k,j}^m\bigg\}
=\lim_{n\to\infty}\mathbf{E}\bigg\{\sum_{j}e^{-uG_{11}}G_{1,j}^m\bigg\},\end{equation*}
and
\begin{equation}
\bigg|\mathbf{E}\bigg\{\sum_{j}e^{-uG_{11}}G_{1,j}^m\bigg\}-V_m(u,z)\bigg|\le Cm(1+u)/n^{1/2}
\label{p1.5}\end{equation}
To average with respect to $a^{(1)}$ we use  the second and the third line of  (\ref{rep_1}) and the formulas:
\begin{equation} \label{p1.6}
R^{-m}=\intd_0^\infty dv\frac{v^{m-1}}{(m-1)!}e^{-Rv},
  \end{equation}
  \begin{equation}\label{p1.9}
    e^{-uR}=1-u^{1/2}\int_0^\infty dv
\frac{\mathcal{J}_1(2\sqrt{uv})}{\sqrt v} \exp\{-R^{-1}v\},
\end{equation}
which are valid for any  $\Re R>0$ and $u\in\mathbb{C}$. Then
we get
\begin{eqnarray}
T_m(u)&:=&\mathbf{E}\bigg\{\sum_{j}e^{-uG_{11}}G_{1,j}^m\bigg\}=\mathbf{E}\mathbf{E}_1\bigg\{\int_0^\infty
dv_1\frac{v_1^{m-1}}{(m-1)!}e^{-v_1(z+(G^{(1)}
a^{(1)},a^{(1)}))}\bigg\}\notag\\
&&-u^{1/2}\mathbf{E}\mathbf{E}_1\bigg\{\int_0^\infty\int_0^\infty dv_1dv_2
\frac{v_1^{m-1}\mathcal{J}_1(2\sqrt{uv_2})}{\sqrt v_2(m-1)!}
e^{-(v_1+v_2)(z+(G^{(1)}
a^{(1)},a^{(1)}))}\bigg\}\notag\\
&&+
\mathbf{E}\mathbf{E}_1\bigg\{\sum_{j>1}(G^{(1)}a^{(1)})_j^m\int_0^\infty
dv_1\frac{v_1^{m-1}}{(m-1)!}e^{-v_1(z+(G^{(1)}
a^{(1)},a^{(1)}))}\bigg\}
\label{p1.7}\\
&&-u^{1/2}\mathbf{E}\mathbf{E}_1\bigg\{\sum_{j>1}(G^{(1)}a^{(1)})_j^m
\cdot\int_0^\infty \int_0^\infty dv_1dv_2\frac{v_1^{m-1}\mathcal{J}_1(2\sqrt{uv_2})}{\sqrt v_2(m-1)!}e^{-(v_1+v_2)(z+(G^{(1)}
a^{(1)},a^{(1)}))}\bigg\}\notag\\
&=&I_{1,m}-u^{1/2}I_{2,m}(u)+I_{3,m}(u)-u^{1/2}I_{4,m}(u).\notag
\end{eqnarray}
Using (\ref{p2.*}) and averaging with respect to $\{a_{1,i}\}$, we have
\begin{eqnarray}
I_{1,m}&=&\int_0^\infty
dv_1\frac{v_1^{m-1}}{(m-1)!}e^{-v_1z}\mathbf{E}\mathbf{E_1}\bigg\{\exp\{-v_1\sum_lG^{(1)}_{ll}a_{1,l}\}
\bigg\}+O(n^{-1})\notag\\
&=&\int_0^\infty
dv_1\frac{v_1^{m-1}}{(m-1)!}e^{-v_1z}\mathbf{E}\prod_l\left(1-\frac{p}{n}+\frac{p}{n}e^{-v_1G^{(1)}_{ll}}\right)
+O(n^{-1})\label{p1.7a}\\
&=&\int_0^\infty
dv_1\frac{v_1^{m-1}}{(m-1)!}e^{-v_1z}\mathbf{E}\{\exp\{-p+pf^{(1)}_n(z,v_1)\}\}(1+O(n^{-1}))+O(n^{-1})\notag\\
&=&\int_0^\infty
dv_1\frac{v_1^{m-1}}{(m-1)!}e^{-v_1z}e^{-p+pf(z, v_1)}+r_{1,m},\notag
\end{eqnarray}
where
\[|r_{1,m}|\le Cmn^{-1/2},\]
and we used  first (\ref{l1.3})--(\ref{ineq*}) to replace $f^{(1)}_n(z,v_1)$ by $f_n(z,v_1)$, and then
(\ref{t3.4}) to replace $f_n(z,v_1)$ by $f(z,v_1)$.
Similarly
\begin{eqnarray}
I_{2,m}&=&\int_0^\infty\int_0^\infty
dv_1dv_2\frac{v_1^{m-1}\mathcal{J}_1(2\sqrt{uv_2})}{\sqrt{v_2}(m-1)!}e^{-z(v_1+v_2)}
e^{-p+ pf(z,v_1+v_2)}+r_{2,m}(u)\label{p1.7b}\\
&&|r_{2,m}(u)|\le Cn^{-1/2}.\notag
\end{eqnarray}
Moreover, using  (\ref{p2.0}) and (\ref{p2.*}), we obtain
\begin{eqnarray}
I_{3,m}&=&\int_0^\infty
dv_1\frac{v_1^{m-1}}{(m-1)!}e^{-v_1z}\mathbf{E}\mathbf{E_1}\bigg\{\sum_{j,k} (G_{jk}^{(1)})^ma_{1k}
\exp\{-v_1\sum_lG^{(1)}_{ll}a_{1,l}\}
\bigg\}+r_{3,m}'\notag\\
&=&\int_0^\infty
dv_1\frac{v_1^{m-1}}{(m-1)!}e^{-v_1z}\mathbf{E}\bigg\{\frac{p}{n}\sum_{j,k} e^{-v_1 G^{(1)}_{kk}}(G_{jk}^{(1)})^m
\exp\{-p+pf^{(1)}_n(z,v_1)\}\bigg\}+r_{3,m}'''\notag\\
&=p&\int_0^\infty
dv_1\frac{v_1^{m-1}}{(m-1)!}e^{-v_1z}e^{-p+pf(z, v_1)}T_m(v_1)+r_{3,m},\label{p1.7c}\\
&&|r_{3,m}(u)|\le Cmn^{-1/2}.\notag
\end{eqnarray}
Here we  used also the relation
\begin{eqnarray*}
 \mathbf{E}\bigg\{\frac{1}{n}\sum_{j,k} e^{-v_1G^{(1)}_{kk}}(G_{jk}^{(1)})^m\bigg\}&=&T_m(v_1)
+O\left(\frac{m}{n}\right).
\end{eqnarray*}
which can be proved similarly to (\ref{p1.2}).
Repeating the argument used for $I_{3,m}$,
we obtain
\begin{eqnarray}
I_{4,m}&=&\int_0^\infty\int_0^\infty
dv_1 dv_2\frac{v_1^{m-1}\mathcal{J}_1(2\sqrt{uv_2})}{\sqrt{v_2}(m-1)!}e^{-z(v_1+v_2)}
e^{-p+pf(z, v_1+v_2)}T_m(v_1+v_2)+r_{3,m},\notag\\
&&|r_{4,m}(u)|^2\le Cmn^{-1/2}.\label{p1.7d}
\end{eqnarray}
Collecting the above relations, we get in view of (\ref{p1.7}) the equation
\begin{eqnarray*}
T_m(u)&=&\varphi_m(u)+\widehat K_m(T_m)(u)+r_m(u),\\
|r_m(u)|&\le& C m(1+\sqrt u)n^{-1/2},
\end{eqnarray*}
where the function $\varphi_m(u)$ is defined by the r.h.s. of (\ref{p1.7a}) and (\ref{p1.7b})
and the integral operator $\widehat K_m$ is defined by the r.h.s. of (\ref{p1.7c}) and (\ref{p1.7d}).
It is easy to see that for $\Re z>2$ the operator norm in the Banach space of the functions
with the norm (\ref{norm}) satisfies the inequality
\[||\widehat K_m||\le q<1.\]
Hence, we get (\ref{p1.5}). Then summing with respect to $m$  and taking into account the bounds
for the remainder terms, we obtain (\ref{p1.1}).
$\square$

\medskip

The next lemma is a technical one. We will use it in the proof of Theorem \ref{t:1}
below.

\begin{lemma}\label{l:2} Set
\begin{equation}\label{D^1}
D^{(1)}(z,u):=n(f_n(z,u)-f^{(1)}_n(z,u))=e^{-uG_{11}}+\sum_{i=2}^{n}\left(e^{-uG_{ii}}
-e^{-uG^{(1)}_{ii}}\right).
\end{equation}
 Then for $\Re z>2$ we have
\begin{eqnarray}\label{s-a.1}
&&\mathbf{Var}\left\{\mathbf{E}_1\left\{e^{-uG_{11}(z)}\right\}\right\}\le n^{-1},\quad
\mathbf{Var}\left\{\mathbf{E}_1\left\{D^{(1)}(z,u)\right\}\right\}\le e^{u}q_1(u)n^{-1},\notag\\
&&\mathbf{Var}\left\{\mathbf{E}_1\left\{e^{-u_1G_{11}(z_1)}D^{(1)}(z_2,u_2)\right\}\right\}\le
e^{u_2}q_2(u_1,u_2)n^{-1}.
\end{eqnarray}
with polynomial $q_1,q_2$. Moreover, if we denote
\begin{eqnarray}\label{V_n}
V_n(z_1,u_1;z_2,u_2)=\mathbf{Cov}_1\left\{e^{-u_1G_{11}(z_1)},D^{(1)}(z_2,u_2)\right\},
\end{eqnarray}
where $\mathbf{Cov}_1\{F_1,F_2\}:=\mathbf{E}_1\{F_1F_2\}-\mathbf{E}_1\{F_1\}\mathbf{E}_1\{F_2\}$,  then there exists
\begin{equation}\label{V}
V(z_1,u_1;z_2,u_2)=\lim_{n\to\infty}V_n(z_1,u_1;z_2,u_2)
\end{equation}
and for any  fixed $z_1,u_1;z_2,u_2$
\begin{equation}\label{conv_V}
|V(z_1,u_1;z_2,u_2)-V_n(z_1,u_1;z_2,u_2)|\le q_3(u_1,u_2)e^{u_2}n^{-1/2},
\end{equation}
with polynomial $q_3$.
\end{lemma}

\textit{Proof of Lemma \ref{l:2}}.
The first bound of (\ref{s-a.1}) can be proved similarly to (\ref{p1.7}) -- (\ref{p1.7c}). Indeed,
according to (\ref{p1.9}) we have
\begin{eqnarray*}
T_0(u)&:=&\mathbf{E}_1\bigg\{e^{-uG_{11}}\bigg\}=1-
u^{1/2}\mathbf{E}_1\bigg\{\int_0^\infty dv
\frac{\mathcal{J}_1(2\sqrt{uv})}{\sqrt v}
e^{-v(z+(G^{(1)}
a^{(1)},a^{(1)}))}\bigg\}.
\end{eqnarray*}
Then, averaging with respect to $\{a_{1,i}\}$ similarly to (\ref{p1.7a}), we get
\[T_0(u)=1-\sqrt{u}\int_0^\infty dv\frac{\mathcal{J}_1(2\sqrt{uv})}{\sqrt v}
e^{-vz}e^{-p+pf(z, v)}+r_{0},\quad \mathbf{E}\{|r_{0}|^2\}\le C/n.
\]
To prove the second bound of (\ref{s-a.1}) we use (\ref{p2.0}), which gives us
that $\mathbf{E}_1\{D^{(1)}-e^{-uG_{11}^{(1)}}\}$ coincides with the r.h.s. of (\ref{p2.0}).
Then  (\ref{p1.9}) for $\widetilde u=i(G^{(1)}_{jk})^2u$ applied to the r.h.s. of (\ref{p2.0})
yields:
\begin{eqnarray*}
\mathbf{E}_1\{D^{(1)}-e^{-uG_{11}^{(1)}}\}&=&
\frac{ip\sqrt u}{n}\int_0^\infty dv\sum_{j,l}e^{-uG^{(1)}_{jj}}G^{(1)}_{jl}
\frac{\mathcal{J}_1\left(2i\sqrt{uv}G^{(1)}_{jl}\right)}{\sqrt v}e^{-zv-p+pf^{(1)}_n(z,v)}
\\
+O(q(u)e^{u}/n^{1/2})&
=&\frac{p}{2}\int_0^\infty dv \,v^{-1}V_{J,n}(z,u,uv)e^{-zv-p+pf(z,v)}+O(e^{cu}/n^{1/2})
\end{eqnarray*}
with $V_{J,n}$ of (\ref{V_F}).
Now the second inequality of (\ref{s-a.1}) follows from Lemma \ref{l:3}, if we use
(\ref{p1.9}) to integrate the bound for $r_{J,n}(z,u,uv)$ of (\ref{p1.1a}) with respect to $v$.
The third bound of (\ref{s-a.1}) follows from the first and the second one.

Relations (\ref{V}) -- (\ref{conv_V}) can be proved if we repeat  the argument
(\ref{p1.7}) -- (\ref{p1.7d}) and then apply Lemma \ref{l:3}.$\square$

\medskip

Now we are ready to prove Theorem \ref{t:1}.

\medskip

\textit{Proof of Theorem \ref{t:1}} Fix  $z_1,\dots,z_m$ such that $\Re z_i\ge 2,\,i=1,\dots m$. We find first $M_{2,n}$.
Using the symmetry of the problem and  Lemma \ref{l:1} it is easy to see that
\begin{eqnarray}
M_{2,n}&=&n\mathbf{E}\left\{\overset{\circ}e^{-u_1G_{11}(z_1)}\overset{\circ}f^{(1)}_n(z_2,u_2)\right\}
+\mathbf{E}\left\{\overset{\circ}e^{-u_1G_{11}(z_1)}
\overset{\circ}D^{(1)}(z_2,u_2)\right\}\notag\\
&=&T_1+V_n(z_1,u_1;z_2,u_2),
\label{t1.6a}\end{eqnarray}
where $V_n(z_1,u_1;z_2,u_2)$ is defined in Lemma \ref{l:2}.
Relations (\ref{rep_1}), (\ref{p1.9}), and   (\ref{p2.*}) yield
\begin{eqnarray}\label{T_1}
T_1&=&-nu_1^{1/2}\intd_0^\infty dv \frac{\mathcal{J}_1(2\sqrt{u_1v})}{\sqrt v}e^{-z_1v}
\mathbf{E}\left\{\overset{\circ}f^{(1)}_n(z_2,u_2)e^{-v(G^{(1)}a^{(1)},a^{(1)})}\right\}\\
&=&-nu_1^{1/2}\intd_0^\infty dv \frac{\mathcal{J}_1(2\sqrt{u_1v})}{\sqrt v}e^{-z_1v}
\mathbf{E}\left\{\overset{\circ}f^{(1)}_n(z_2,u_2)\left(\prod_{k}e^{-vG^{(1)}_{kk}a_{1k}}+
r_v\right)\right\}
\notag\end{eqnarray}
with $r_v$ of (\ref{p2.*})

 Since $\overset{\circ}f^{(1)}_n(z_2,u_2)$ does not depend
on $\{a_{1j}\}_{j=2}^n$ we can average with respect to $a^{(1)}$ and similarly to
(\ref{p2.**})  obtain
\begin{eqnarray*}
\left|\mathbf{E}_1\{ r_v\}\right|\le C(v+v^2)/n.
\end{eqnarray*}
We  used that (\ref{p2.a}) and the first bound of
(\ref{f<1}) for $||G^{(1)}||$.
 The bound, the Schwarz inequality, and Lemma \ref{l:1}
 yield
\begin{eqnarray*}
\bigg|\mathbf{E}\bigg\{\overset{\circ}f^{(1)}_n(z_2,u_2)r_v\bigg\}\bigg|\le n^{-1}|\Re
z|^{-2}\mathbf{E}\bigg\{|\overset{\circ}f^{(1)}_n(z_2,u_2)|\bigg\}\le C(v+v^2)/n^{3/2}.
\end{eqnarray*}

 Then, integrating with respect to $v$ (recall that $|\mathcal{J}_1|\le
1$) and averaging $\prod_{k}e^{-vG^{(1)}_{kk}a_{1k}}$ over $\{a_{1k}\}$, we get similarly to
(\ref{p1.7a}):
\begin{eqnarray*}
T_1&=&-nu_1^{1/2}\intd_0^\infty dv \frac{\mathcal{J}_1(2\sqrt{u_1v})}{\sqrt v}e^{-z_1v-p}
\mathbf{E}\left\{\overset{\circ}f^{(1)}_n(z_2,u_2)\left(\prod_{k}e^{-vG^{(1)}_{kk}a_{1k}}\right)\right\}
+O((u/n)^{1/2})\\
&=&-nu_1^{1/2}\intd_0^\infty dv\frac{\mathcal{J}_1(2\sqrt{u_1v})}{\sqrt v}e^{-z_1v-p}
\mathbf{E}\left\{\overset{\circ}f^{(1)}_n(z_2,u_2)e^{pf_n^{(1)}(z_1,v)}\right\}+O((u/n)^{1/2}).
 \notag\end{eqnarray*}
Writing
$f_n^{(1)}(z_1,v)=\mathbf{E}\{f_n^{(1)}(z_1,v)\}+\overset{\circ}f^{(1)}_n(z_1,v)$, we have
\begin{eqnarray}
T_1&=&-nu_1^{1/2}\intd_0^\infty dv\frac{\mathcal{J}_1(2\sqrt{u_1v})}{\sqrt v}e^{-z_1v-p}
\mathbf{E}\left\{\overset{\circ}f^{(1)}_n(z_2,u_2)e^{p\mathbf{E}\{f_n^{(1)}(z_1,v)\}
+p\overset{\circ}f_n^{(1)}(z_1,v)}\right\}+r_{n}^{(2)}\notag\\
&=&n\intd_0^\infty dvK_n(u_1,v;z_1)\mathbf{E}\left\{\overset{\circ}f^{(1)}_n(z_2,u_2)
\bigg(p\overset{\circ}f_n^{(1)}(z_1,v)+
O\bigg((\overset{\circ}f_n^{(1)}(z_1,v))^2\bigg)\bigg)\right\}+r_{n}^{(2)}\notag\\
&=&\intd_0^\infty dvK_n(u_1,v;z_1)
M_{2,n}(z_1,v;z_2,u_2)+r_{n}^{(3)},
\label{int_eq}\end{eqnarray}
 where
\begin{eqnarray}
K_n(u_1,v;z_1)&:=-&u_1^{1/2}\frac{\mathcal{J}_1(2\sqrt{u_1v})}{\sqrt v}e^{-z_1v-p}e^{
\mathbf{E}\{pf_{n}^{(1)}(v,z)\}}\notag\\
r_{n}^{(2)}&=&O((u_1/n)^{1/2})
\notag\\
r_{n}^{(3)}&=&r_{n}^{(2)}+n\intd_0^\infty dv K_n(u_1,v;z_1)
\mathbf{E}\bigg\{\overset{\circ}f_{n}^{(1)}(z_2,u_2)O\bigg((\overset{\circ}f_n^{(1)}(z_1,v))^2\bigg)\bigg\}\notag\\
&\le& Cn^{-1/2}u_1^{1/2}.
\label{K}\end{eqnarray}
The last bound follows from (\ref{M^1-M}).

Thus, we obtain that
\begin{eqnarray}\label{equ}
M_{2,n}(z_1,u_1;z_2,u_2)&=&\intd_0^\infty dv K_n(u_1,v;z_1)
M_{2,n}^{(1)}(z_1,v;z_2,u_2)+\notag\\
&&V_n(z_1,u_1;z_2,u_2)+r_{n}^{(3)}(z_1,u_1;z_2,u_2)+O(n^{-1/2}),
\end{eqnarray}
where $V_n(z_1,u_1;z_2,u_2)$ is defined in (\ref{V_n}).
Besides,  using (\ref{t3.1}) and the inequality $|\mathcal{J}_1(x)|\le 1$,
we obtain that   uniformly in $u,v>0$
\[\lim_{n\to\infty}K_n(u,v;z)=-u^{1/2}\frac{\mathcal{J}_1(2\sqrt{uv})}{\sqrt v}e^{-zv-p}\exp\{
-\mathbf{E}\{pf(z,v)\}\}=:K(u,v;z),\quad\Re z>2,\]
and
\[|K_n(u,v;z)-K(u,v;z)|\le Cuv^{-1/2}e^{-|\Re z|v}n^{-1/2}.\]
 Using the above bounds
to replace  $K_n$ by $K$ in (\ref{equ}) and (\ref{conv_V})
to replace $V_n$ by $V$, we can write (\ref{equ}) in the form
\begin{eqnarray}\label{equ1}
M_{2,n}(z_1,u_1;z_2,u_2)&=&\intd_0^\infty dv K(u_1,v;z_1)
M_{2,n}^{(1)}(z_1,v;z_2,u_2)+\notag\\
&&V(z_1,u_1;z_2,u_2)+r_{n}^{(4)}(z_1,u_1;z_2,u_2)+O(n^{-1/2}),\\
&&|r_{n}^{(4)}(z_1,u_1;z_2,u_2)|\le q(u_1,u_2)e^{u_2}n^{-1/2}
\notag\end{eqnarray}
with polynomial $q$. The inequality
\begin{equation}\label{b_K}|K(u,v;z)|\le u^{1/2}v^{-1/2}e^{-\Re zv}\end{equation}
implies that there exists $M_0>2$ such that for all $z$ with $\Re z>M_0$ the norm of
the integral operator $K$ in the
Banach space $\mathcal{H}$ (see (\ref{norm})) satisfy the inequality
\begin{equation}\label{||K||}
||K||\le \frac{1}{2}.
\end{equation}
and so there exists the inverse operator $(I-K)^{-1}$. But the problem is that
the bound for $r_{n}^{(4)}$ above does not allow us to conclude that
 $r_{n}^{(4)}\in\mathcal{H}$ (recall that we fixe $u_2$ and consider $r_{n}^{(4)}$ as a function
 of $u_1$). This difficulty can be easily overcome if
we consider a new function
\[\widetilde {M}_{2,n}(z_1,u_1;z_2,u_2)=M_{2,n}(z_1,u_1;z_2,u_2)-r_{n}^{(4)}(z_1,u_1;z_2,u_2).\]
 Then (\ref{equ1}) takes the form
\begin{eqnarray}\label{equ2}
\widetilde M_{2,n}(z_1,u_1;z_2,u_2)&=&\intd_0^\infty dv K(u_1,v;z_1)
\widetilde M_{2,n}^{(1)}(z_1,v;z_2,u_2)+\notag\\
&&V(z_1,u_1;z_2,u_2)+K(r_{n}^{(4)})(z_1,u_1;z_2,u_2)+O(n^{-1/2}),
\end{eqnarray}
 and (\ref{b_K}) yields
\[ |K(r_{n}^{(4)})(z_1,u_1;z_2,u_2)|\le C\sqrt u n^{-1/2}. \]
Thus we can apply the $(I-K)^{-1}$ to (\ref{equ2}) and obtain that for any $z:\Re z>M_0$ there exists
 the limit
\begin{equation}\label{M_2}
M_{2}(z_1,u_1;z_2,u_2):=\int_0^{\infty}(I-K)^{-1}(u_1,v;z_1)V(z_1,v;z_2,u_2) \,d v.
\end{equation}
But according to Lemma \ref{l:1} $M_{2,n}(z_1,u_1;z_2,u_2)$ is an analytic function bounded uniformly
 in each compact in the right half plane of $\mathbb{C}$. Hence, taking any bounded domain $U$ which
contains some $z$: $\Re z>M_0$, for any fixed $u_1,u_2$ we can choose a subsequence $M_{2,n_k}(z_1,u_1;z_2,u_2)$
which converges uniformly in $z_1\in U$ to some analytic in $U$ function. But since for $z$: $\Re z>M_0$
for any convergent subsequence there exists a unique limit of $M_{2,n_k}(z_1,u_1;z_2,u_2)$,
defined by (\ref{M_2}), on the basis of the uniqueness theorem we conclude that for any $z\in U$
there exists a limit of $M_{2,n}(z_1,u_1;z_2,u_2)$ and this limit for $\Re z>M_0$ is defined by (\ref{M_2}).
Hence we have proved (\ref{t1.1})  for $m=2$.

For arbitrary $m$
we have instead of (\ref{t1.6a})
\begin{eqnarray}
M_{m,n}&:=&n^{m/2+1/2}\mathbf{E}\left\{\overset{\circ}e^{-u_1G_{11}(z_1)}\right.
\prod_{j=2}^m\left.\left(n^{-1}D^{(1)}(z_j,u_j)+\overset{\circ}f^{(1)}_n(z_j,u_j)\right)\right\}+O(n^{-1/2})
\notag\\
&=&n^{m/2+1/2}\mathbf{E}\left\{\overset{\circ}e^{-u_1G_{11}(z_1)}\prod_{j=2}^m
\overset{\circ}f^{(1)}_n(z_j,u_j)\right\}\notag\\&&+
\sum_{j=2}^mn^{(m-1)/2}\mathbf{E}\left\{\overset{\circ}e^{-u_1G_{11}(z_1)}D^{(1)}(z_j,u_j)
\prod_{i\not=j}\overset{\circ}f^{(1)}_n(z_i,u_i)\right\}\notag\\
&&+O(n^{-1/2})=:T_1+\sum_{j=2}^mT_{2j}+O(n^{-1/2})
\label{t1.6}\end{eqnarray}
Then, similarly to (\ref{int_eq}), we write $T_1$ from the r.h.s. of (\ref{t1.6})
as
\begin{eqnarray}
T_1&=&\intd_0^\infty dv K_n(u_1,v;z_1)
M_{m,n}(z_1,v;\dots;z_m,u_m)\notag\\
&&+r_{n}^{(3)}(z_1,u_1;\dots;z_m,u_m)+O(n^{-1/2}q(u_1,\dots,u_m)),\label{t1.15}\end{eqnarray}
where $r_{n}^{(3)}$ admits the bound (\ref{K}).

Since  $f_{n}^{(1)}$ does not depend on $\{a_{1j}\}_{j=2}^n$ we can average with respect to these
variables and, using (\ref{s-a.1}) write $T_{2j}$ in the form
\begin{eqnarray}\label{t1.16}
T_{2j}&=&\mathbf{E}\bigg\{\left(\mathbf{E}_1\{e^{-u_1G_{11}(z_1)}D^{(1)}(z_j,u_j)\}
-\mathbf{E}_1\{e^{-u_1G_{11}(z_1)}\}\mathbf{E}_1\{D^{(1)}(z_j,u_j)\}\right)\notag\\
&&
\cdot\prod_{i=2,i\not=j}^m\overset{\circ}f_{n}^{(1)}(z_j,u_j)\bigg\}=
\mathbf{E}\{V_n(z_1,u_1;z_j,u_j)\}\mathbf{E}\bigg\{\prod_{i=2,i\not=j}^m
\overset{\circ}f_{n}^{(1)}(z_j,u_j)\bigg\}\notag\\
&&
+\mathbf{E}\bigg\{\overset\circ V_n(z_1,u_1;z_j,u_j)\prod_{i=2,i\not=j}^m
\overset{\circ}f_{n}^{(1)}(z_j,u_j)\bigg\}
\end{eqnarray}
Using the Schwartz inequality and Lemmas 1,2, it is easy to obtain that the last term in the r.h.s.
of (\ref{t1.16}) is $O(n^{-1/2})$. Hence, (\ref{t1.6}), (\ref{t1.15}) and (\ref{t1.16}) yield
\begin{eqnarray}
M_{m,n}(z_1,u_1;\dots;z_m,u_m)=\intd_0^\infty dv K_n(u_1,v;z_1)
M_{m,n}(z_1,v;\dots;z_m,u_m)\notag\\
+\sum_{j=1}^m\mathbf{E}\{V_n(z_1,u_1;z_j,u_j)\}
M_{m-2,n}(z_2,v_2;\dots;z_{j-1},u_{j-1};z_{j+1},u_{j+1},\dots;z_m,u_m)\notag\\
+O(n^{-1/2}q_1(u_1,\dots,u_m)(e^{u_2}+\dots+e^{u_j})),\label{t1.15}\end{eqnarray}
Then, using once more the argument, which we applied to (\ref{M_2}), we can prove (\ref{t1.1}) first for
$\Re z>2$ and then extend it to the whole right half plane of $\mathbb{C}$.
$\square$

\medskip

\textit{Proof of Theorem \ref{t:4}}
We prove Theorem \ref{t:4} in two steps: first for polynomial $\varphi$
and then  extend the statement to any
 real valued functions $\varphi$, satisfying conditions of the theorem.
 For polynomial $\varphi$
 we replace in Theorem \ref{t:2} the product of
traces of resolvent of $A$ with different $z_j$ (see (\ref{M^*})) by the
product of traces of  $\varphi_1(A),\dots,\varphi_p(A)$ with $\varphi_1,\dots,\varphi_m$ being
some fixed polynomials. More precisely,
we consider (cf (\ref{M^*}))
\[M_{p,n}(\varphi_1,\dots,\varphi_m):=
n^{-m/2}
\mathbf{E}\bigg\{\hbox{Tr }\overset\circ{\varphi_1}(A)\dots\hbox{Tr }\overset\circ{\varphi_m}(A)\bigg\}=
\mathbf{E}\bigg\{\prod_{j=1}^m n^{-1/2}
\overset{\circ}{\mathcal{N}}_n[\varphi_j]\bigg\}\]
and prove that for any $m$ and any fixed polynomial $\varphi_1,\dots,\varphi_m$ there exists  the limit
\begin{equation}\label{lim}
\lim_{n\to\infty}M_{m,n}(\varphi_1,\dots,\varphi_m)=
M_{m}(\varphi_1,\dots,\varphi_m)\end{equation}
and
\begin{equation}\label{wick}
M_{m}(\varphi_1,\dots,\varphi_m)=\sum_{j=2}^m M_2(\varphi_1,\varphi_j)M_{m-2}(\varphi_2,\dots,\varphi_{j-1},
\varphi_{j+1},\dots,\varphi_m).
\end{equation}
Then taking
$\varphi_1=\dots=\varphi_m=P$ we obtain that there exist the limits of all moments
of $ n^{-1/2}\overset{\circ}{\mathcal{N}}_n[P]$ and these moments are expressed in terms of the
second moment by the same way as for the Gaussian random variable.

Recall that Theorem \ref{t:2} imply that the (\ref{lim}) and (\ref{wick}) are valid for
$\varphi_{z_j}(\lambda)=(i\lambda-z_j)^{-1}$. We will replace $\varphi_{z_j}$ by the
polynomial $\varphi_{j}$ in (\ref{lim}) -- (\ref{wick}) step by step, starting from the
last one $\varphi_{z_m}(\lambda)$. To this end we
prove by induction  with respect to the polynomial degree $k$  that  if we replace $\varphi_{z_m}(\lambda)$ by a polynomial $P_k(\lambda)$
of degree not exceeding  $k$, then (\ref{lim}) -- (\ref{wick})
 are valid.

 For $k=0,1$ $\overset{\circ}{\mathcal{N}}_n[P_k]=0$ (recall that
$A_{jj}=0$), so  (\ref{lim}) -- (\ref{wick}) are trivial. Let us assume that that we know
(\ref{lim}) -- (\ref{wick}) for $\varphi_m(\lambda)=P_l(\lambda)$ with $l\le k-1$
and prove that they are valid for $l=k$. Consider
\begin{eqnarray}\label{lim_phi}
\varphi_{m}(\lambda)&=&\varphi(\lambda,z_m,k)=-z_m\lambda^{k}(i\lambda-z_m)^{-1}\\&=&
-(-i)^{k}z_m\left(z^{k}_m(i\lambda-z_m)^{-1}+\sum_{l=1}^{k}C_{k}^l(i\lambda-z_m)^{l-1}z_m^{k-l}\right).
\notag\end{eqnarray}
By the above representation and
the induction assumption (\ref{lim}) and (\ref{wick}) are valid for
$\varphi_{m}(\lambda)=\varphi(\lambda,z_m,k)$ with any $z_m$.
Moreover, if we use the inequalities
\begin{eqnarray}\label{b_pol}
    \mathbf{E}\bigg\{\bigg|n^{-1/2}\overset{\circ}{\mathcal{N}}_n[P_k^*]\bigg|^m\bigg\}&\le &C(m,k),
    \quad P_k^*(\lambda)=\lambda^{k},\\
    \mathbf{E}\bigg\{\bigg|n^{-1/2}\overset{\circ}{\mathcal{N}}_n[\varphi_k^*]\bigg|^m\bigg\}&\le&
    C(m,k)/|\Re z|,
    \quad\varphi^*_k(\lambda)=\lambda^{k}(i\lambda-z)^{-1},\quad k,m\in \mathbb{N},\notag
\end{eqnarray}
combined with the H\"{o}lder inequality
\[|M_{m,n}(\varphi_1,\dots,\varphi_m)|\le
\prod_{j=1}^m\mathbf{E}^{1/m}\left\{|n^{-1/2}\overset{\circ}{\mathcal{N}}_n[\varphi_j]|^m\right\},\]
then,  since
$P_{k}^*(\lambda)-\varphi(\lambda;z_m,k)=-i\varphi^*_{k+1}(\lambda)$, we obtain
\begin{equation}\label{uni_b}
|M_{m,n}(\varphi_1,\dots,P_{k}^*)-M_{m,n}(\varphi_1,\dots,\varphi(.\,;z_m,k))|=
|M_{m,n}(\varphi_1,\dots,\varphi^*_{k+1})|\le
\frac{C}{|\Re z_m|^{1/m}},
\end{equation}
 where $C$ does not depend on $n$ and $z_m$.
We will prove (\ref{b_pol}) later.
Now let us use a simple proposition
\begin{proposition}\label{p:*} Let the sequence of the functions $\{u_n(\zeta)\}_{n=1}^\infty$ converges point-wise
to the function $u(z)$, as $n\to\infty$,
in the domain $\Re\zeta>C$, and for any fixed $n$ $u_n(\zeta)\to u_n^*$, as
$\Re\zeta\to\infty$, so that
\begin{equation}\label{conv}
|u_n(\zeta)-u_n^*|\le C_0/|\Re\zeta|^\alpha,\quad \alpha>0.
\end{equation}
Then there exist the limits
\begin{equation}\label{*}
\lim_{n\to\infty}u_n^*=\lim_{\Re\zeta\to\infty}u(\zeta)
\end{equation}
\end{proposition}
\textit{Proof.} Take any $\varepsilon>0$ and $\zeta_\varepsilon$ such that
$C_0/|\Re\zeta_\varepsilon|^\alpha\le \varepsilon/4$. Moreover, choose $N$ such that
$|u_n(\zeta_\varepsilon)-u(\zeta_\varepsilon)|\le\varepsilon/4$ for any $n\ge N$.
Then for any $n,n'>N$
\[|u_n^*-u_{n'}^*|\le
|u_n^*-u_n(\zeta_\varepsilon)|+|u_{n'}^*-u_{n'}(\zeta_\varepsilon)|+|u_n(\zeta_\varepsilon)
-u(\zeta_\varepsilon)|+|u_{n'}(\zeta_\varepsilon)
-u(\zeta_\varepsilon)|\le \varepsilon.\]
Hence, there exists $u^*=\lim_{n\to\infty}u_n^*$. In addition, for any  $\zeta$ and any $\varepsilon>0$
one can choose $N$ such that
$|u_N(\zeta)-u(\zeta)|\le\varepsilon/2$ and $|u_N^*-u^*|\le\varepsilon/2$. Then
\[ |u(\zeta)-u^*|\le |u(\zeta)-u_N(\zeta)|+|u_N(\zeta)-u_N^*|+|u_N^*-u^*|\le\varepsilon+
C_0/|\Re\zeta|^\alpha.
\]
Thus, there exists the second limit in (\ref{*}) and it coincides with $u^*$.$\square$

Now if  for fixed $z_1,\dots,z_{m-1}$ we consider the functions
$u_n(z_m)=M_{m,n}(\varphi_1,\dots,\varphi(.\,;z_m,k))$, then  (\ref{lim_phi}) gives
the point-wise convergence of $u_n(z_m)$ and
(\ref{uni_b}) coincides with (\ref{conv})  of Proposition \ref{p:*} with
$u_n^*=M_{k,n}(\varphi_1,\dots,P_k^*)$. Applying the proposition we obtain that
(\ref{lim}) -- (\ref{wick})  are valid if we replace the last function $\varphi_m$ by any
polynomial of degree $k$.

  Repeating the above procedure we  replace step by step all $\varphi_1,\dots\varphi_{m-1}$ by
  polynomials of any fixed degree.  As it was mentioned about this implies that for any polynomial
  $P$   $n^{-1/2}\overset{\circ}{\mathcal{N}}_n[P]$
converges in distribution to a gaussian random variable with zero
mean and the variance from (\ref{CLT_pol}). Hence, by  the
standard argument we conclude that uniformly in $x$ varying in any
compact of $\mathbb{R}$ \begin{equation}\label{CLT_pol}
\mathbf{E}\left\{e^{ixn^{-1/2}\overset{\circ}{\mathcal{N}}_n[P]}\right\}=e^{-x^2/2V(P)},\quad
V(P)=\lim_{n\to\infty}\mathbf{Var}\{n^{-1/2}\mathcal{N}_n[P]\}.
\end{equation}
To finish the proof of CLT for polynomials  we are
left to prove (\ref{b_pol}). It is done in the further proof of
Theorem \ref{t:4}.

To extend CLT to a  wider class of functions we use
\begin{proposition} \label{p:CLTcont}
Let $\{\xi_{l}^{(n)}\}_{l=1}^{n}$ be a triangular array of random variables,
$\displaystyle\mathcal{N}_{n}[\varphi ]=\sum_{l=1}^{n}\varphi
(\xi _{l}^{(n)})$ be its linear statistics,
corresponding to a test function $\varphi :\mathbb{R}\rightarrow
\mathbb{R}$, and
\[V_{n}[\varphi]=\mathbf{Var}\{n^{-1/2}\mathcal{N}_{n}[\varphi ]\}\]
be the variance of $\mathcal{N}_{n}[\varphi ]$. Assume that

(a) there exists a vector space $\mathcal{L}$ endowed with a norm $||...||$
and such that $V_{n}$ is defined on $\mathcal{L}$ and admits the bound
\begin{equation}
V_{n}[\varphi ]\leq C||\varphi ||^2,\;\forall \varphi \in \mathcal{L},
\label{ocVn1}
\end{equation}%
where $C$ does not depend on $n$;

(b) there exists a dense linear manifold $\mathcal{L}_{1}\subset
\mathcal{L}$ such that the Central Limit Theorem is valid for
$\mathcal{N}_{n}[\varphi ],\;\varphi \in \mathcal{L}_{1}$, i.e., if%
\begin{equation*} Z_{n}[x\varphi ]=\mathbf{E}\left\{
e^{ixn^{-1/2}\overset{\circ}{\mathcal{N}}_{n}[\varphi ]}\right\}
\end{equation*}%
is the characteristic function of
$n^{-1/2}\overset{\circ }{\mathcal{N}}_{n}[\varphi ]$, then there
exists a continuous quadratic functional $V:\mathcal{L}_{1}\rightarrow \mathbb{R}_{+}$
such that we have uniformly in $x$,
varying on any compact interval
\begin{equation}\label{limZ}
\lim_{n\rightarrow \infty }Z_{n}[x\varphi ]=e^{-x^{2}V[\varphi
]/2},\;\forall \varphi \in \mathcal{L}_{1};
\end{equation}
Then  $V$ admits a continuous extension to $\mathcal{L}$ and Central Limit
Theorem is valid for all $\mathcal{N}_{n}[\varphi ]$,
 $\varphi \in \mathcal{L}$.
\end{proposition}

\textit{Proof.} Let $\{\varphi _{k}\}$ be a sequence of elements of
$\mathcal{L}_{1}$ converging to $\varphi \in \mathcal{L}$. We have
then in view of the inequality $ |e^{ia}-e^{ib}|\leq |a-b|$, the
linearity of $\overset{\circ}{\mathcal{N}} _{n}[\varphi ]$ in
$\varphi $, the Schwarz inequality, and (\ref{ocVn1}):
\begin{eqnarray*} \Big|Z_{n}(x\varphi
)-Z_{n}(x\varphi)|_{\varphi=\varphi_k} \Big| &\leq
&|x|\mathbf{E}\left\{ \left\vert
n^{-1/2}\overset{\circ}{\mathcal{N}}_{n}[\varphi
]-n^{-1/2}\overset{\circ}{ \mathcal{N}}_{n}[\varphi _{k}]\right\vert
\right\} \\ &\leq
&|x|\mathbf{Var}^{1/2}\{n^{-1/2}\mathcal{N}_{n}[\varphi -\varphi
_{k}]\}\leq C|x|\;||\varphi -\varphi _{k}||.
\end{eqnarray*}
Now,
passing first to the limit $n\rightarrow \infty $ and then
$k\rightarrow \infty $, we obtain the assertion. $\square$

\medskip

Let us show now that hypothesis (a) and (b) of Proposition \ref{p:CLTcont} are fulfilled
in some vector space.
We fix some $c>0$ and consider the vector space $\mathcal{L}$ of functions $\varphi$
 such that $\varphi,\,\varphi',\,\varphi''\in
L^2(\mathbb{R},\cosh^{-2}(c\lambda))$ (see (\ref{L_2})).  Denote
\[||\varphi||^2=\int|\varphi''(\lambda)|^2\cosh^{-2}(c\lambda)d\lambda+
\int|\varphi'(\lambda)|^2\cosh^{-2}(c\lambda)d\lambda+
\int|\varphi(\lambda)|^2\cosh^{-2}(c\lambda)d\lambda\]
It is evident that the space of all
polynomials $\mathcal{L_1}$ is  dense subspace in $\mathcal{L}$ with respect to the norm $||.||$.
 Moreover, (\ref{CLT_pol})
proves (b). Hence we are left to check assumption (a) of Proposition \ref{p:CLTcont}.

It is easy to see
that if $\varphi\in\mathcal{L}$ then
$f(\lambda)=\varphi(\lambda)\cosh^{-1}(c\lambda)\in L_2(\mathbb{R})$
and also  $f',f''\in L_2(\mathbb{R})$ and
\[||f||_{L_2(\mathbb{R})}^2+||f''||_{L_2(\mathbb{R})}^2\le
C||\varphi||^2.\] Hence it is enough to check that
\begin{equation}\label{t4.1}
  \textbf{Var}\{n^{-1/2}\hbox{Tr } f(A)e^{\pm cA}\}\le
  C(||f||_{L_2(\mathbb{R})}^2+||f''||_{L_2(\mathbb{R})}^2).
\end{equation}
According to Proposition \ref{p:mart} (see (\ref{ppcg}))
\begin{equation}\label{t4.2}
\textbf{E}\left\{\left|n^{-1/2}\overset{\circ}{\hbox{Tr }} f(A)e^{\pm cA}\right|^{2m}\right\}
\le C_{2m}\mathbf{E}\left\{\left|\hbox{Tr }
 (f(A)e^{\pm cA}-f(A^{(1)})e^{\pm cA^{(1)}})\right|^{2m}\right\},
\end{equation}
where $A^{(1)}$ is defined in (\ref{A^1}). Note that to prove (\ref{t4.1}) it suffices to consider
$m=1$, but we need other $m$ to prove (\ref{b_pol}). Write
\[\hbox{Tr }\left(f(A)e^{ cA}-f(A^{(1)})e^{ cA^{(1)}}\right)=\int d\xi\widehat f(\xi)
\hbox{Tr }\left(e^{(i\xi+c)A}-e^{(i\xi+c)A^{(1)}}\right),\]
where $\widehat f$ is the Fourier transform of $f$. Then the Duhamel formula yields
\begin{eqnarray*}
\bigg|\hbox{Tr }\left(e^{(i\xi+c)A}-e^{(i\xi+c)A^{(1)}}\right)\bigg|&=&\bigg|\int_0^1dt(i\xi+c)\hbox{Tr }
\left(e^{t(i\xi+c)A}(A-A^{(1)})e^{(i\xi+c)A^{(1)}(1-t)}\right)\bigg|\\
&=&\bigg|\int_0^1dt(i\xi+c)\sum_{j=1}^n(e^{t(i\xi+c)A})_{j1}(e^{(i\xi+c)A^{(1)}(1-t)}a^{(1)})_j\bigg|\\
&\le&(e^{2tcA}e_1,e_1)^{1/2}e^{2cA^{(1)}(1-t)}a^{(1)},a^{(1)})^{1/2}.
\end{eqnarray*}
Here we used that
\begin{eqnarray*}\hbox{Tr }
\left(e^{t(i\xi+c)A}(A-A^{(1)})e^{(i\xi+c)A^{(1)}(1-t)}\right)=
\sum_{j,k}(e^{t(i\xi+c)A})_{j1}a_{1k}(e^{(i\xi+c)A^{(1)}(1-t)})_{kj}\\+
\sum_{j,k}(e^{t(i\xi+c)A})_{jk}a_{k1}(e^{(i\xi+c)A^{(1)}(1-t)})_{1j}.
\end{eqnarray*}
 The first some gives
$\sum_{j=1}^n(e^{t(i\xi+c)A}e_1)_{j}(e^{(i\xi+c)A^{(1)}(1-t)}a^{(1)})_j$
where $e_1=(1,0,\dots,0)$ and the vector $a^{(1)}$ is defined in
(\ref{rep_1}). The second sum is 0, since relations $A^{(1)}_{i1}=A^{(1)}_{1i}=0$
($i=1,\dots,n$) imply  $(e^{(i\xi+c)A^{(1)}(1-t)})_{1i}=0$ ($i=1,\dots,n$). Then the Schwarz inequality
yields
\begin{eqnarray}\notag \textbf{Var}\{n^{-1/2}\hbox{Tr }f(A)e^{
cA}\}&\le& C_2\bigg(\int|\widehat f(\xi)|(|\xi|+c)d\xi\bigg)^2
\mathbf{E}^{1/2}\{(e^{2tcA}e_1,e_1)^2\}\\&&\mathbf{E}^{1/2}\{(e^{2(1-t)cA^{(1)}}a^{(1)},a^{(1)})^2\}.
\label{t4.3}\end{eqnarray}
Using the Schwarz inequality once more
and then the symmetry of the problem, we obtain
\[\mathbf{E}\{(e^{2tcA}e_1,e_1)^2\}\le\mathbf{E}\{(e^{4tcA}e_1,e_1)\}=
\mathbf{E}\{\hbox{Tr }e^{4tcA}\}.
\]
 Similarly, using the Schwarz
inequality and then the independence $A^{(1)}$ of $a^{(1)}$, we can
average with respect to $a^{(1)}$ to obtain
\begin{eqnarray*}\mathbf{E}\{(e^{2(1-t)cA^{(1)}}a^{(1)},a^{(1)})^2\}\le
\mathbf{E}\{(e^{4(1-t)cA^{(1)}}a^{(1)},a^{(1)})(a^{(1)},a^{(1)})\}\\
\le C(p+p^2) \mathbf{E}\{n^{-1}\hbox{Tr
}e^{4(1-t)cA^{(1)}}\}+C(p^2+p^3)\mathbf{E}^{1/2}\{n^{-1}\hbox{Tr
}e^{8(1-t)cA^{(1)}}\}.
 \end{eqnarray*}
 Since all entries of $A$ and
$A^{(1)}$ and $A-A^{(1)}$ are positive, we have for any $t$
\[\mathbf{E}\{\hbox{Tr }e^{4(1-t)cA^{(1)}}\}\le\mathbf{E}\{\hbox{Tr
}e^{4(1-t)cA}\}\le\mathbf{E}\{\hbox{Tr }e^{4cA}\}.
\]
Moreover,
according to the result of \cite{BG2} we have for any $m$
\[\mathbf{E}\{n^{-1}\hbox{Tr }A^{2m}\}\le
C^m_0m!\quad\Rightarrow\quad \mathbf{E}\{n^{-1}\hbox{Tr }e^{cA}\}\le
2e^{C_0c^2/2}.
\]
In addition the Schwarz inequality yields
\[\bigg(\int|\widehat f(\xi)|(|\xi|+c)d\xi\bigg)^{4}\le\int|\widehat
f(\xi)|^2(|\xi|+c)^4d\xi \int(|\xi|+c)^{-2}d\xi\le
C(||f||^2_{L^2(\mathbb{R})}+||f''||^2_{L^2(\mathbb{R})}).
\]
Summarizing  the above inequalities,  we obtain (\ref{t4.1}) and
hence the assumption (a) of Proposition \ref{p:CLTcont}.
Then Theorem \ref{t:4} follows from Proposition \ref{p:CLTcont}.

To prove (\ref{b_pol}) we use again (\ref{t4.2}), where for the first line of (\ref{t4.2})
$f(\lambda)=\lambda^k\cosh^{-1}(c\lambda)$ and for the second line
$f(\lambda)=\lambda^k\cosh^{-1}(c\lambda)$. Repeating the above argument we obtain (\ref{t4.2}).
$\square$

\end{document}